\documentclass[%
 reprint,
superscriptaddress,
 amsmath,amssymb,
 aps,
pre,
]{revtex4-1}

\usepackage[english]{babel}
\usepackage{amsfonts}
\usepackage{amsmath}
\usepackage{amssymb}
\usepackage{graphicx,psfrag}
\usepackage{color}
\usepackage{epigraph}
\usepackage{pxfonts}
\usepackage{stmaryrd}
\usepackage{listings}
\usepackage[pdftitle={Continuum Percolation Thresholds},
pdfauthor={Stephan Mertens and Cristopher Moore},
pdfcreator={},
pdfsubject={Percolation},
pdfproducer={},
pdfkeywords={Monte Carlo, percolation, union-find},hyperfootnotes=false]{hyperref}
\usepackage{url}
\IfFileExists{date.tex}{%
  \input{date.tex}

}

\newcommand*{\e}{\mathrm{e}}

\newcommand{\find}{\texttt{find}}
\newcommand{\union}{\texttt{merge}}

\begin{document}

\title{Continuum Percolation Thresholds in Two Dimensions}

\author{Stephan Mertens}
\email{mertens@ovgu.de}
\affiliation{Santa Fe Institute, 1399 Hyde Park Rd., Santa Fe, NM 87501, USA}
\affiliation{Institut f\"ur Theoretische Physik, Universit\"at
  Magdeburg, Universit\"atsplatz~2, 39016~Magdeburg, Germany}

\author{Cristopher Moore}
\email{moore@santafe.edu}
\affiliation{Santa Fe Institute, 1399 Hyde Park Rd., Santa Fe, NM 87501, USA}

\date{\today}

\begin{abstract}
A wide variety of methods have been used to compute percolation thresholds. 
In lattice percolation, the most powerful of these methods consists of microcanonical simulations 
using the union-find algorithm to efficiently determine the connected clusters, 
and (in two dimensions) using exact values from conformal field theory for the probability, at the phase transition, 
that various kinds of wrapping clusters exist on the torus.  
We apply this approach to percolation in continuum models, finding overlaps between objects with real-valued positions and orientations.  
In particular, we find precise values of the percolation transition for disks,
squares, rotated squares, and rotated sticks in two dimensions, and confirm that these transitions behave as conformal field theory predicts.  
The running time and memory use of our algorithm are essentially linear as a function of the number of objects at criticality.
\end{abstract}


\maketitle


\section{Introduction}
\label{sec:intro}

For more than 50 years, percolation theory has been used to model
static and dynamic properties of porous media and other disordered
physical systems \cite{gilbert:61,stauffer:aharony,sahimi}. Most natural systems
correspond to continuum percolation, yet most analytical and
numerical work has focused on lattice percolation. This is
reasonable since continuum and lattice percolation lie in the same
universality class.  
For properties that are non-universal, however, such as the location of the threshold, 
one has to study discrete and continuum models individually, and it is 
also satisfying to confirm universality experimentally by measuring critical exponents and 
crossing probabilities.

\begin{figure}
  \centering
  \includegraphics[width=\columnwidth,clip=true]{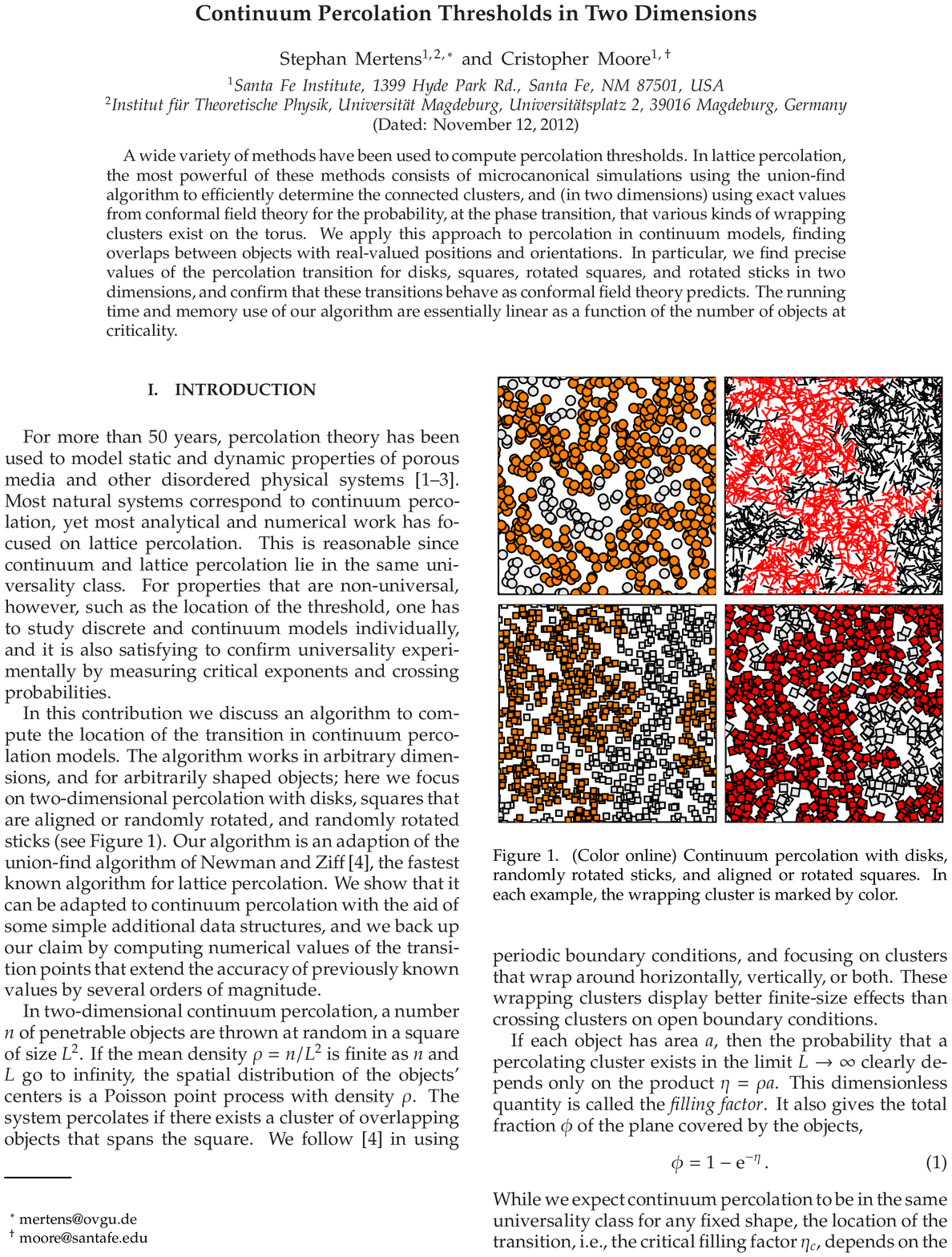}
  \caption{(Color online) Continuum percolation with disks, randomly rotated sticks, and aligned or rotated squares. 
  In each example, the wrapping cluster is marked by color. }
  \label{fig:examples}
\end{figure}

In this contribution we discuss an algorithm to compute the location
of the transition in continuum percolation models. The algorithm works
in arbitrary dimensions, and for arbitrarily shaped objects; here we focus on 
two-dimensional percolation with disks, squares that are aligned or randomly rotated, 
and randomly rotated sticks (see Figure~\ref{fig:examples}).  
Our algorithm is an adaption of the union-find algorithm of Newman and Ziff~\cite{newman:ziff:01}, 
the fastest known algorithm for lattice percolation.  We show that it can be adapted to continuum percolation 
with the aid of some simple 
additional data structures, and we back up our claim by computing numerical
values of the transition points that extend the accuracy of previously
known values by several orders of magnitude.

In two-dimensional continuum percolation, a number $n$ of penetrable
objects are thrown at random in a square of size $L^2$.
If the mean density $\rho=n/L^2$ is finite as $n$ and $L$ go to
infinity,  the spatial distribution of the objects' centers is a
Poisson point process with density $\rho$.  The system percolates 
if there exists a cluster of overlapping objects that spans the square.  
We follow~\cite{newman:ziff:01} in using periodic boundary conditions, 
and focusing on clusters that wrap around horizontally, vertically, or both.  
These wrapping clusters display better finite-size effects than crossing clusters 
on open boundary conditions.

If each object has area $a$, then the probability that a percolating cluster exists 
in the limit $L \to \infty$ clearly depends only on the product $\eta = \rho a$.  
This dimensionless quantity is called the \emph{filling factor}.  
It also gives the total fraction $\phi$ of the plane covered by the objects, 
\begin{equation}
  \label{eq:2}
  \phi = 1 - \e^{-\eta} \, . 
\end{equation}
While we expect continuum percolation to be in the same universality class for 
any fixed shape, the location of the transition, i.e., the critical filling factor $\eta_c$,
depends on the shape of the objects.  We write $\eta_c^\medcirc$, $\eta_c^\Box$, 
$\eta_c^\Diamond$, and $\eta_c^\times$ for the percolation of disks, aligned squares, 
randomly rotated squares, and randomly rotated sticks.  In defining $\eta$, we treat 
sticks of length $\ell$ as if they have area $a=\ell^2$.

\begin{table*}
  \centering
  \begin{tabular}{r|l|l|l|l}
     &  $\eta_c^\medcirc$ & $\eta_c^\Box$ & $\eta_c^\Diamond$ & $\eta_c^\times$ \\ \hline 
    previous & 
    1.128085(2)  & 
    1.0982(3) & 
    0.9819(6) &
    5.63726(2) \\ 
    our work & 
    1.12808737(6) & 
    1.09884280(9) &
    0.9822723(1) & 
    5.6372858(6)
  \end{tabular}
  
    \begin{tabular}{r|l|l|l|l}
     &  $\phi_c^\medcirc$ & $\phi_c^\Box$ & $\phi_c^\Diamond$ & $\phi_c^\times$ \\ \hline 
    previous & 
    0.6763475(6) &
    0.6665(1) & 
    0.6254(2) &
    0.99643738(7) \\ 
    our work & 
    0.67634831(2) & 
    0.66674349(3) & 
    0.62554075(4) & 
    0.996437475(2)
  \end{tabular}
  \caption{Numerical values of critical filling factors $\eta_c$ and area factors $\phi_c=1-\e^{-\eta_c}$ in continuum percolation 
  for disks, aligned squares, randomly rotated squares, and randomly rotated sticks.  
   Previous estimates are from~\cite{quintanilla:ziff:07,baker:etal:02,torquato:jiao:12,li:zhang:09}.}
  \label{tab:etas}
\end{table*}

Table~\ref{tab:etas} lists the most accurate numerical values for
$\eta_c$ from previous work and the work presented here.  
The best previous results on disk percolation are due to Quintanilla, Torquato, and Ziff~\cite{quintanilla:ziff:07} 
who varied the density of the Poisson process as a function of position and kept track of the front of the connected cluster.  
The best previous results on aligned squares are due to Torquato and Jaio~\cite{torquato:jiao:12}, who rescale an initial set of particles 
so that its density is close to rigorous bounds.  
The best previous results on rotated squares are due to Baker et al.~\cite{baker:etal:02}.  
The best previous results on sticks are due to Li and Zhang~\cite{li:zhang:09}, who used an approach 
similar to ours but with open boundary conditions.

Our results are consistent with the rigorous bounds 
\begin{eqnarray}
  1.127 &\,\leq\, &\eta_c^\medcirc \,\leq\, 1.12875 \nonumber \\
  1.098 &\,\leq\, &\eta_c^\Box \,\leq\, 1.0995 \, , \label{eq:balister}
\end{eqnarray}
computed with $99.99\%$ confidence by Balister, Bollob\'as and Walters~\cite{balister:05} 
using a Monte Carlo estimate of a high-dimensional integral.  On the other hand, it is a little sad to dash the 
hope---which one might have entertained after reading~\cite{baker:etal:02,torquato:jiao:12}, and which is just barely 
consistent with~\eqref{eq:balister}---that $\phi_c^\Box$ is exactly $2/3$.

In the following sections, we review the union-find algorithm
of~\cite{newman:ziff:01}, how it finds wrapping clusters in periodic
boundary conditions, and how we extend it to the continuous case.  We
show that the running time of our algorithm is essentially linear in
the number of objects, i.e., linear in $L^2$.  In addition to
estimating the threshold, we also measure the finite-size exponent
$\nu$, giving strong evidence that these continuum models are in the
same universality class as lattice percolation.  Finally, we find that
the probability of a wrapping cluster at criticality is precisely that
predicted by conformal field theory.

\section{The Algorithm}

We will simulate percolation in the microcanonical ensemble, i.e., where the number $n$ of objects in the square is 
fixed.  In each trial, we add one object at a time, stopping as soon as a percolating cluster appears.  
Following~\cite{newman:ziff:01}, we keep track of the connected components at each step using the union-find data structure.  
In union-find, each cluster is represented uniquely by one of its members.  We have access to two functions: 
$\find(i)$, which finds the representative $r(i)$ of the cluster to which object $i$ belongs, and $\union(i,j)$, which merges 
$i$'s cluster and $j$'s cluster together into a single one with the same representative.  

Internally, union-find works in a very simple way.  Each object $i$ is linked to a unique ``parent'' $p(i)$ in the same cluster, 
except for the representative which has no parent.  When we call $\find(i)$, it follows the links from $i$ to its parent $p(i)$, its grandparent $p(p(i))$, 
and so on, until it reaches $i$'s representative $r(i)$.  Similarly, $\union(i,j)$ uses $\find(i)$ and $\find(j)$ to obtain $r(i)$ and $r(j)$, 
and declares one of them to be the parent of the other, unless $r(i)=r(j)$ and they are already in the same cluster. 

The running time of $\find(i)$ is proportional to the length of the path from $i$ to $r(i)$.  If $\union(i,j)$ sensibly links the smaller cluster 
to the larger one, setting $p(r(i))=r(j)$ whenever $i$'s cluster is smaller than $j$'s, a simple inductive argument shows that  
these paths never exceed $\log_2 n$ in length.  
However, we can make these paths even shorter using a trick called \emph{path compression}.  Since $r(i)$ is the representative of every object $j$
along the path from $i$ to $r(i)$, we can set $p(j)=r(i)$ for all of them, linking them directly to their representative so that $\find$ will work 
in a single step the next time we call it.

As a result, the \emph{amortized cost} of the $\find$ and $\union$
operations---that is, the average cost per operation over the course
of many operations---is nearly constant.  Specifically, it is
proportional to $\alpha(n)$, when $\alpha$ is the inverse of the
Ackermann function~\cite{tarjan:75}.  The Ackermann function grows
faster than any primitive recursive function, i.e., any function that
can be computed with a fixed number of for-loops: faster than an
exponential, an iterated tower of exponentials, and so on~\cite{noc}.
As a consequence, $\alpha(n)$ grows incredibly slowly, and the
smallest value of $n$ such that $\alpha(n) > 4$ is so large that it
can only be written with exotic notation.  Thus the total running time
for $n$ objects is essentially $O(n)$.

\begin{figure}
\begin{center}
\includegraphics[width=0.3\columnwidth]{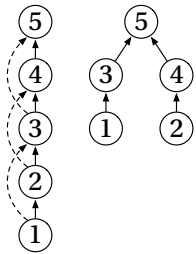}
\end{center}
\caption{When we call $\find(i)$, we split and shorten the path from
  $i$ to its representative $r(i)$ by setting the parent of each object along the path 
  to be its grandparent.  This turns a path of length $\ell$ into two
  paths of length $\ell/2$.}
\label{fig:path-splitting}
\end{figure}

In our implementation, we employ a form of path compression that is
faster and almost as effective: we link each object $j$ on the path to
its grandparent, setting $p(j)=p(p(j))$.  This is known as \emph{path
  splitting}, since it turns a path of length $\ell$ into two paths of
length $\ell/2$, or $(\ell+1)/2$ and $(\ell-1)/2$ if $\ell$ is odd, as
shown in Figure~\ref{fig:path-splitting}.  It has the advantage of
requiring only one pass along the path, and it takes just one line of
code (e.g.~\cite[Appendix A]{newman:ziff:01}).  Like path compression, it guarantees an amortized running time
of $O(\alpha(n))$~\cite{tarjan:van-leeuwen:84}.

\begin{figure}
\begin{center}
\includegraphics[width=0.9\columnwidth]{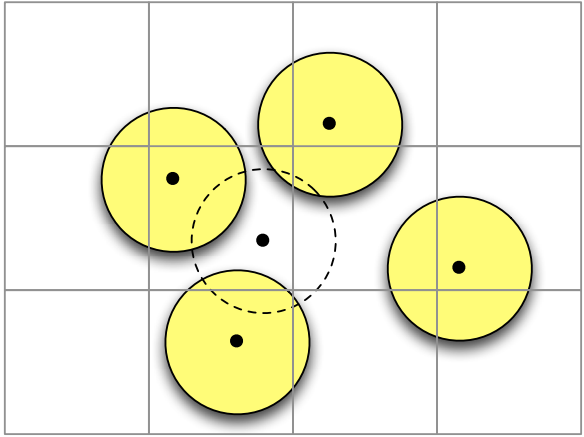}
\end{center}
\caption{(Color online) We divide the plane into square bins whose width equals the diameter of the disks.  Each disk in a given bin (dashed) can only intersect 
with other disks in the same bin, or in the eight neighboring bins.}
\label{fig:bins}
\end{figure}

For lattice percolation as in~\cite{newman:ziff:01}, each time we add a new occupied site, we can check which of its neighbors are occupied, 
and $\union$ them together with the new site.  
In the continuous case, we have more work to do: if we add a new disk (say), we have to find which nearby disks it intersects.  
To do this efficiently, we divide the plane into square bins as shown in Figure~\ref{fig:bins}.  Each disk belongs to whichever bin its center lies in.  
The width of each bin is the diameter of the disks, so that a disk in a given bin can only intersect with other disks in that bin or the eight bins in its 
neighborhood.  

On average, the number of disks in each bin is a constant proportional to $\rho$, so we can find all the disks intersecting with each  
new one in constant time.  We use the same approach for the other shapes; for rotated squares of width $\ell$, the bins need to have 
width $\sqrt{2} \ell$.  A similar approach for rotated sticks was used in~\cite{li:zhang:09}.

If we wished to detect crossing clusters---those that connect, say,
the top and bottom edges of the square---we could add two special
objects to the union-find data structure, which are connected by fiat
to all the disks in the bins along the top or bottom edge.  We would
then check, at each step, whether these two objects are in the same
cluster.  However, as discussed below and in~\cite{newman:ziff:01},
the finite-size scaling is much better if we use periodic boundary
conditions instead, and look for clusters that wrap around the torus
horizontally or vertically.

We detect these wrapping clusters using a technique originally used
for detecting crossing clusters in the Potts
model~\cite{machta:etal:96}.  We associate a vector with each object
in the union-find data structure, recording the displacement between
it and its parent.  In principle this displacement is real-valued, but
it suffices to record an integer vector giving the displacement
between their respective bins.  When we compress and splint a path, we
sum these vectors to get the total displacement between each object on
the path and its new parent.

Now suppose that $\union(i,j)$ finds that two overlapping disks $i$
and $j$ are already in the same cluster.  Object $i$ now has two paths
to its representative; one that goes through its own parent, and
another that consists of hopping to $j$ and then going through $j$'s
parents.  We sum the displacement vectors along both these paths.  If
these sums are the same, then the cluster is simply-connected.  But if
they differ by $\pm L$ in either coordinate, then the cluster has a
nontrivial winding number around one or both directions on the torus.

\begin{figure}
  \centering
  {
    \includegraphics[width=\columnwidth,clip=true]{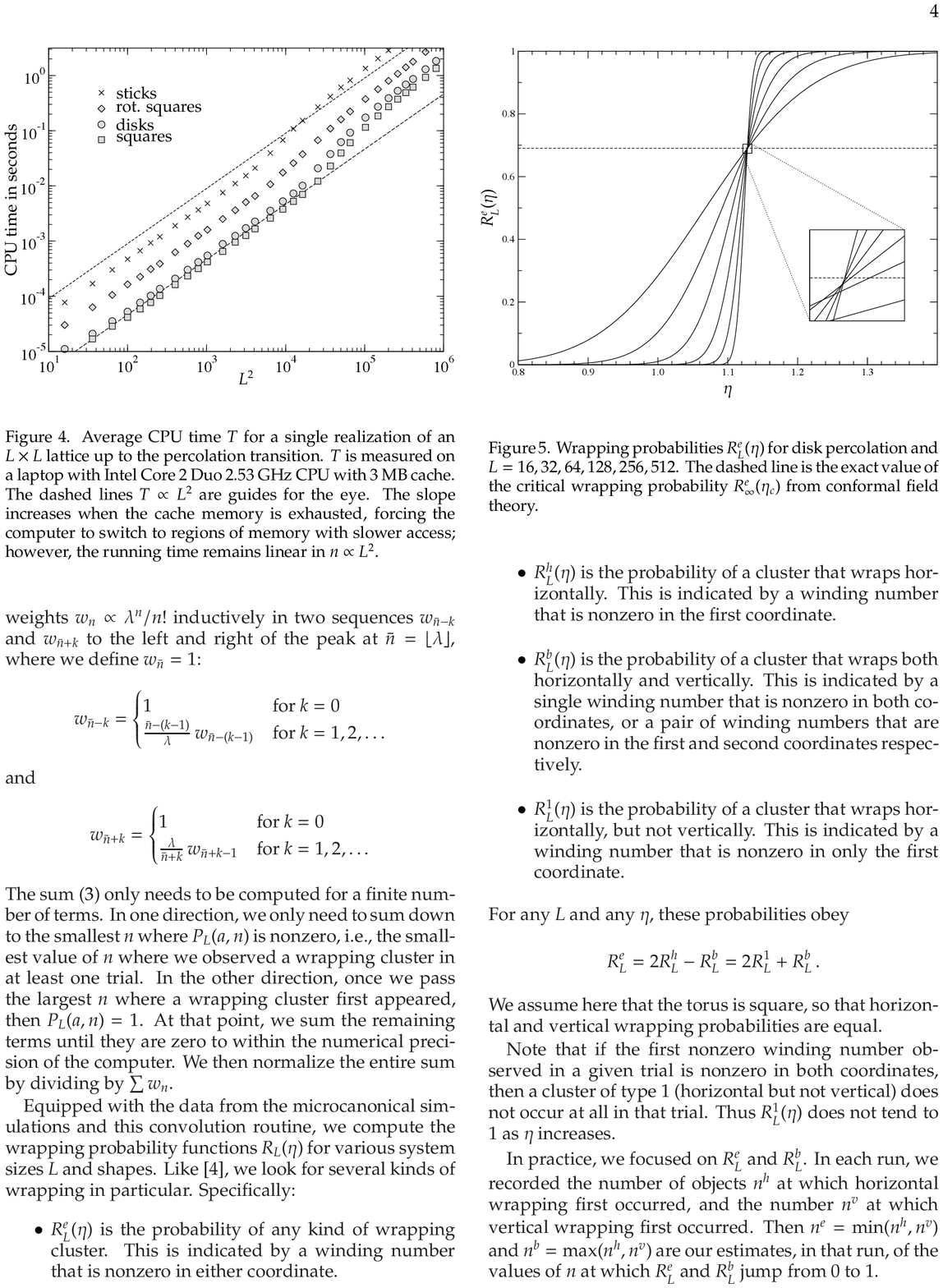}
  }
  \caption{Average CPU time $T$ for a single realization of an $L \times L$
    lattice up to the percolation transition. $T$ is measured on a laptop with Intel
    Core 2 Duo 2.53 GHz CPU with 3 MB cache. The dashed lines 
    $T \propto L^2$ are guides for the eye.  The slope increases when the 
    cache memory is exhausted, forcing the computer to switch to regions 
    of memory with slower access; however, the running time remains linear in $n \propto L^2$.}
  \label{fig:times}
\end{figure}

Like the union-find algorithm itself, the time it takes to sum these
vectors is proportional to the length of the paths from $i$ and $j$ to
their representative.  As Figure~\ref{fig:times} shows, the total
running time of our entire algorithm---the time it takes to carry out
a trial on a lattice of size $L$, adding objects one at a time until a
wrapping cluster appears---is essentially linear in the number $n$ of
objects at criticality, or equivalently linear in $L^2$.  It slows
down somewhat when the computer is forced to switch to parts of its
memory with slower access, but this only affects the leading constant.

\section{Analysis and Results}

If in each trial we stop at the first $n$ where a wrapping cluster
appears, then the estimated probability $P_L(a,n)$ that a wrapping
cluster exists in the microcanonical ensemble with $n$ objects of area
$a$ is the fraction of trials that stop on or before the $n$th step.
To obtain the probability $R_L(\eta)$ of percolation in the grand
canonical ensemble with filling fraction $\eta$, we convolve $P_L$
with the Poisson distribution with mean $\lambda = \rho L^2 = \eta L^2 / a$:
\begin{equation}
  \label{eq:convolution}
  R_L(\eta) = \e^{-\lambda} 
  \sum_{n=0}^\infty \frac{\lambda^n}{n!} \,P_L(a,n) \, . 
\end{equation}
To avoid numerical difficulties where the numerator and denominator are both very large, 
we compute Poisson weights $w_n \propto \lambda^n/n!$ inductively in two sequences
$w_{\bar{n}-k}$ and $w_{\bar{n}+k}$ to the left and right of the peak 
at $\bar{n}=\lfloor \lambda \rfloor$, where we define $w_{\bar{n}}=1$:
\[
  w_{\bar{n}-k} = \begin{cases}
      1 & \text{for $k=0$} \\
      \frac{\bar{n}-(k-1)}{\lambda}\, w_{\bar{n}-(k-1)} & \text{for $k=1,2,\ldots$}
  \end{cases}
\]
and
\[
  w_{\bar{n}+k} = \begin{cases}
      1 & \text{for $k=0$} \\
      \frac {\lambda}{\bar{n}+k} \, w_{\bar{n}+k-1}& \text{for $k=1,2,\ldots$}
  \end{cases}
\]
The sum~\eqref{eq:convolution} only needs to be computed for a finite number of terms. 
In one direction, we only need to sum down to the smallest $n$ where $P_L(a,n)$ is nonzero, 
i.e., the smallest value of $n$ where we observed a wrapping cluster in at least one trial.  In the other direction, 
once we pass the largest $n$ where a wrapping cluster first appeared, then $P_L(a,n)=1$.  
At that point, we sum the remaining terms until they are zero to within the numerical precision
of the computer.  We then normalize the entire sum by dividing by $\sum w_n$.  

Equipped with the data from the microcanonical simulations and this 
convolution routine, we compute the wrapping probability
functions $R_L(\eta)$ for various system sizes $L$ and shapes.  
Like~\cite{newman:ziff:01}, we look for several kinds of wrapping in
particular.  Specifically:
\begin{itemize}
\item $R^e_L(\eta)$ is the probability of any kind of wrapping
  cluster.  This is indicated by a winding number that is nonzero in
  either coordinate.
\item $R^h_L(\eta)$ is the probability of a cluster that wraps
  horizontally.  This is indicated by a winding number that is nonzero in the first coordinate.
\item $R^b_L(\eta)$ is the probability of a cluster that wraps both
  horizontally and vertically.  This is indicated by a single winding number
  that is nonzero in both coordinates, or a pair of winding numbers that are nonzero 
  in the first and second coordinates respectively.  
\item $R^1_L(\eta)$ is the probability of a cluster that wraps horizontally, 
  but not vertically.  This is indicated by a
  winding number that is nonzero in only the first coordinate.
\end{itemize}
For any $L$ and any $\eta$, these probabilities obey 
\[
R_L^e = 2 R_L^h - R_L^b = 2 R_L^1 + R_L^b \, . 
\]
We assume here that the torus is square, so that horizontal and vertical wrapping probabilities are equal.  

Note that if the first nonzero winding number observed in
a given trial is nonzero in both coordinates, then a cluster of type $1$ (horizontal but not vertical) 
does not occur at all in that trial.  Thus $R^1_L(\eta)$ does not tend to $1$
as $\eta$ increases.  

In practice, we focused on $R_L^e$ and $R_L^b$.  In each run,  
we recorded the number of objects $n^h$ at which horizontal wrapping first occurred, 
and the number $n^v$ at which vertical wrapping first occurred.  Then $n^e = \min(n^h,n^v)$ and 
$n^b = \max(n^h,n^v)$ are our estimates, in that run, of the values of $n$ at which $R^e_L$ and $R^b_L$ 
jump from $0$ to $1$.

\begin{figure}
  \centering
  {
    \includegraphics[width=\columnwidth,clip=true]{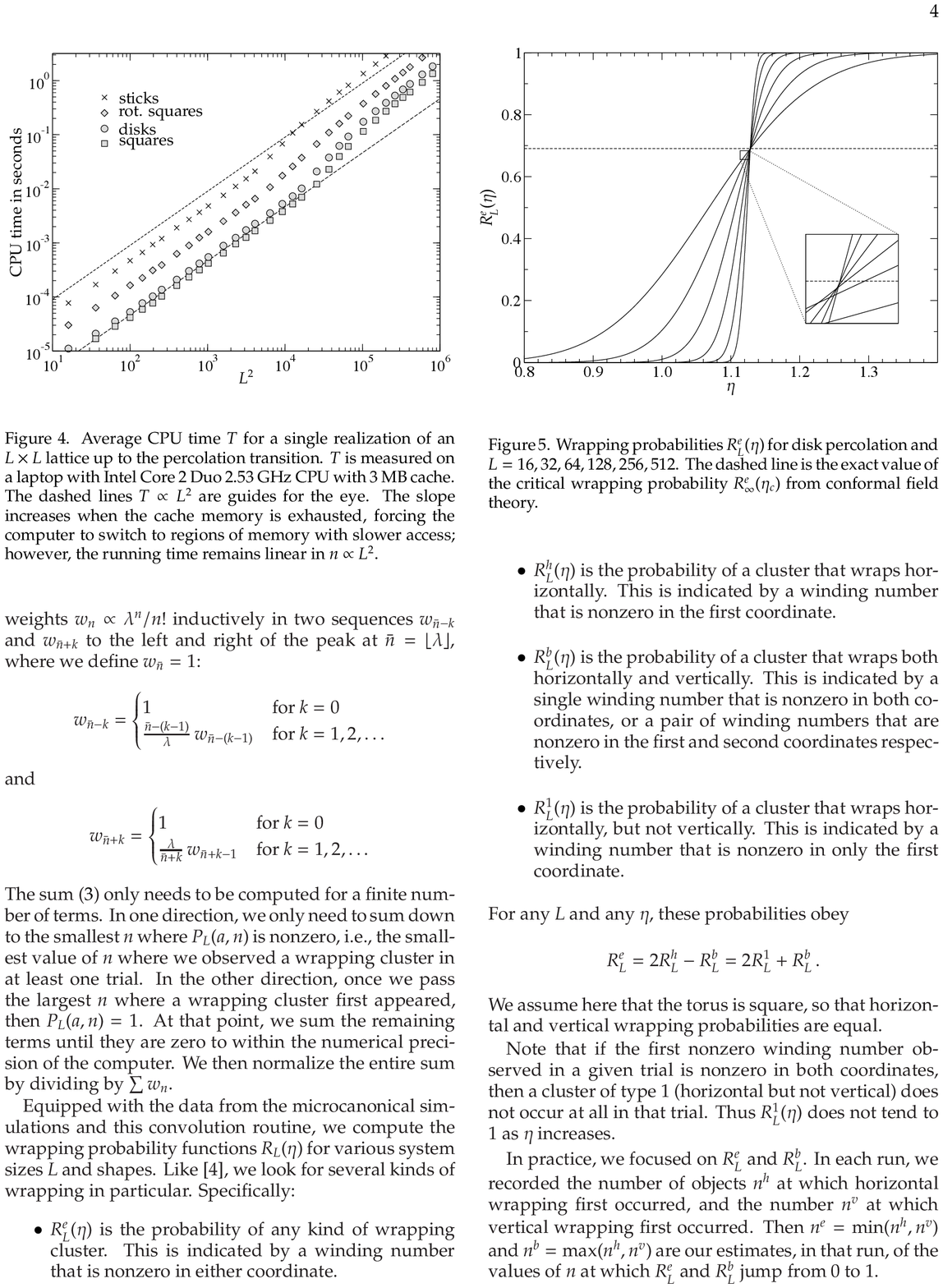}
  }
  \caption{Wrapping probabilities $R^e_L(\eta)$ for disk percolation
    and $L=16,32,64,128,256,512$. The dashed line is the 
    exact value of the critical wrapping probability $R_\infty^e(\eta_c)$ from conformal field theory.}
  \label{fig:span}
\end{figure}

\pagebreak 

A beautiful fact is that, even though the percolation threshold
$\eta_c$ is not known for any of our models, conformal field theory
implies exact values for these probabilities at the transition in the
limit $L \to \infty$~\cite{pinson:94,newman:ziff:01}.  Specifically,
\begin{equation}
  \label{eq:cft-values}
  \begin{aligned}
     R_\infty^h  &= 0.521\,058\,289\,248\,821\,787\,848... \\
     R_\infty^e &= 0.690\,473\,724\,570\,168\,677\,230... \\
     R_\infty^b &= 0.351\,642\,853\,927\,474\,898\,465... \\
     R_\infty^1 &= 0.169\,415\,435\,321\,346\,889\,383... 
  \end{aligned}
\end{equation}
For each $L$, and each type of wrapping cluster, we can estimate
the critical filling factor $\eta_L$ as the solution of the equation
\begin{equation}
  \label{eq:estimator}
  R_L(\eta_L) = R_\infty \, .
\end{equation}
For instance, Figure~\ref{fig:span} shows $R^e_L(\eta)$ for disks for $L$ ranging up to $512$.  
The filling factors $\eta_L$ where these curves cross $R_\infty^e$ rapidly converge to $\eta_c$.  

\begin{figure}
  \centering
  {
    \includegraphics[width=\columnwidth,clip=true]{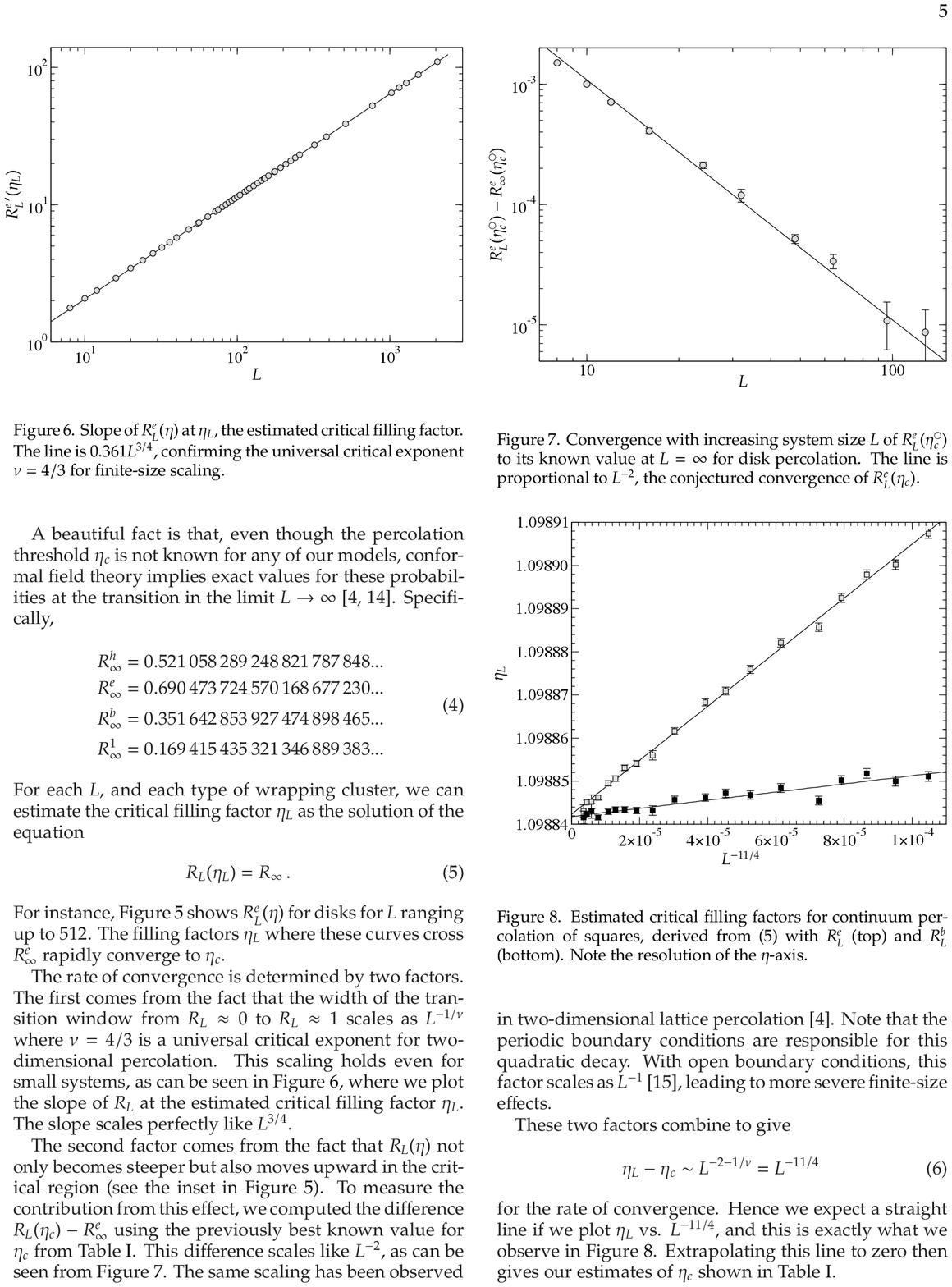}
  }
  \caption{Slope of $R^e_L(\eta)$ at $\eta_L$, the estimated critical
    filling factor. The line is $0.361 L^{3/4}$, confirming the universal critical exponent 
    $\nu=4/3$ for finite-size scaling.}
  \label{fig:steigung}
\end{figure}

The rate of convergence is determined by two factors. The first comes from the fact that the
width of the transition window from $R_L \approx 0$ to $R_L \approx 1$ scales 
as $L^{-1/\nu}$ where $\nu=4/3$ is a universal critical
exponent for two-dimensional percolation.  This scaling holds even for small systems, 
as can be seen in Figure~\ref{fig:steigung}, where we plot the slope of $R_L$
at the estimated critical filling factor $\eta_L$. The slope scales
perfectly like $L^{3/4}$.

\begin{figure}
  \centering
  {
    \includegraphics[width=\columnwidth,clip=true]{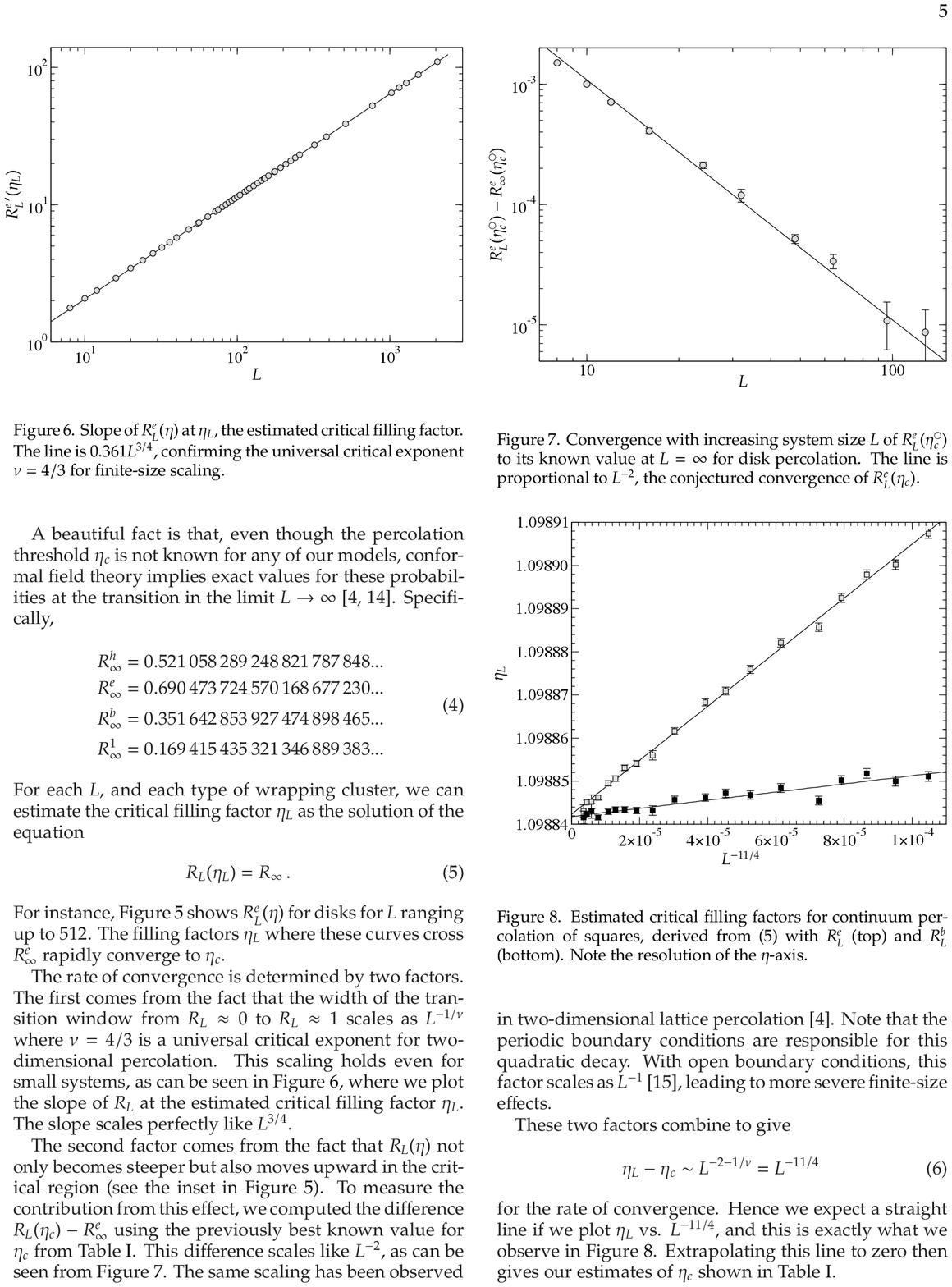}
  }
  \caption{Convergence with increasing system size $L$ of
    $R^e_L(\eta_c^\medcirc)$ to its known value at $L=\infty$ for disk
    percolation. The line is proportional to $L^{-2}$, the
    conjectured convergence of $R^e_L(\eta_c)$.}
  \label{fig:disks-fss}
\end{figure}

The second factor comes from the fact that $R_L(\eta)$ not only
becomes steeper but also moves upward in the critical region (see the
inset in Figure \ref{fig:span}). To measure the contribution from this
effect, we computed the difference $R_L(\eta_c) - R^e_\infty$ using 
the previously best known value for $\eta_c$ from
Table~\ref{tab:etas}.  This difference scales like $L^{-2}$, as can be
seen from Figure~\ref{fig:disks-fss}.  The exponent $-2$ correponds to
the leading irrelevant renormalization exponent $y_i$ in the Kac table
\cite{hu:bloete:deng:13}.  
Note that the periodic boundary conditions are responsible for this decay.  
With open boundary conditions, this factor scales as $L^{-1}$~\cite{hovi:aharony:96}, 
leading to more severe finite-size effects.  

\begin{figure}
  \centering
  {
    \includegraphics[clip=true, width=\columnwidth]{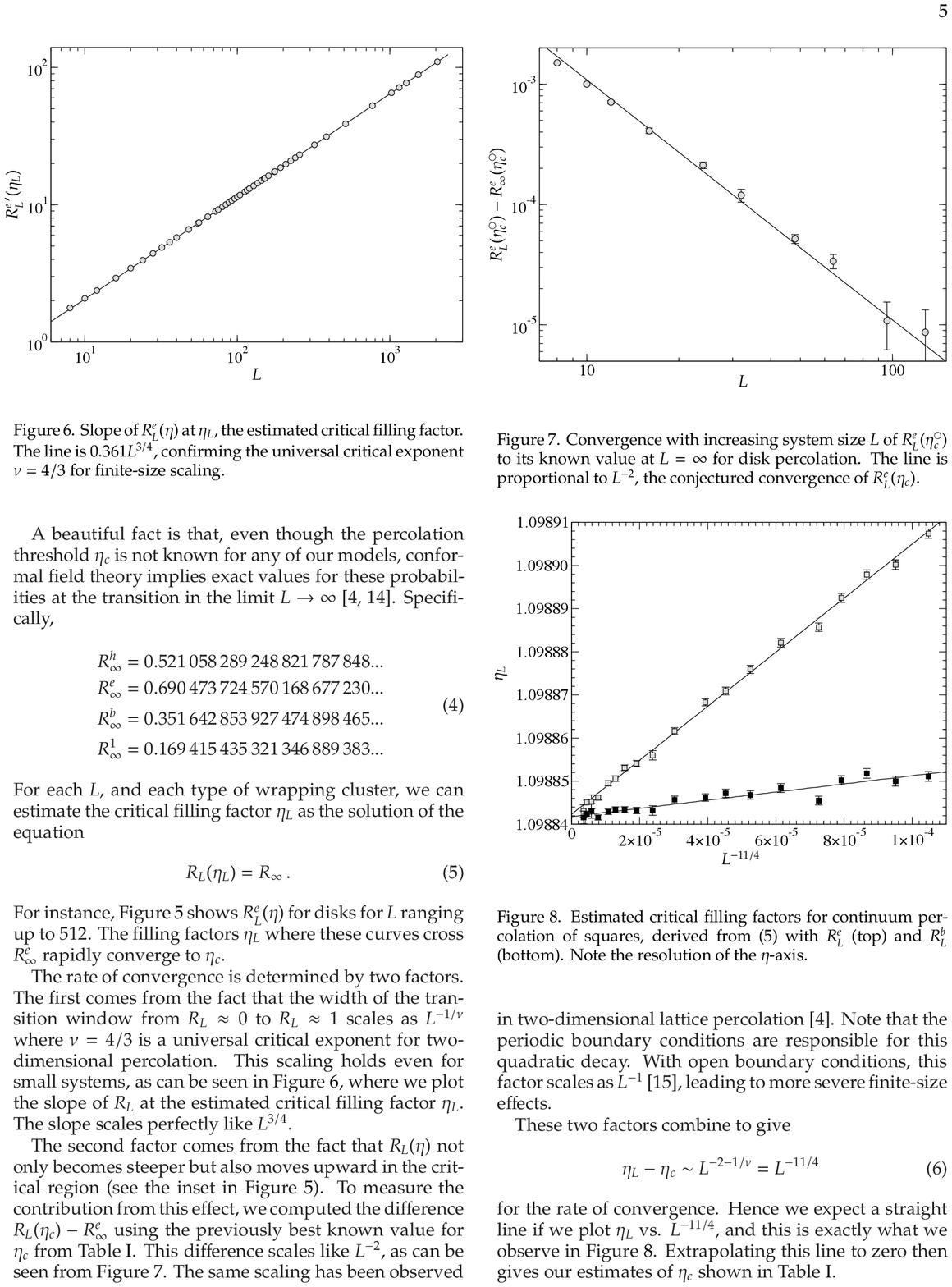}
  }
  \caption{Estimated critical filling factors for continuum
    percolation of squares, derived from \eqref{eq:estimator} with 
    $R_L^e$ (top) and $R_L^b$ (bottom). Note the resolution of the $\eta$-axis.}
  \label{fig:etac}
\end{figure}

These two factors combine to give 
\begin{equation}
  \label{eq:estimator-convergence}
  \eta_L-\eta_c \sim L^{-2-1/\nu} = L^{-11/4}\,
\end{equation}
for the rate of convergence. Hence we expect a straight line if we
plot $\eta_L$ vs. $L^{-11/4}$, and this is exactly what we observe in 
Figure~\ref{fig:etac}. Extrapolating this line to zero then gives our
estimates of $\eta_c$ shown in Table~\ref{tab:etas}.

\pagebreak
How do we compute the error bars in our estimates of $\eta_c$?  
First consider the fluctuations in $R_L(\eta)$.  Each of our microcanonical experiments 
contributes to our estimate of $R_L(\eta)$ for all $\eta$ through the convolution~\eqref{eq:convolution}.   
We can imagine this as choosing $n$ from the Poisson distribution, adding $n$ objects, and returning 
an estimate of $R_L(\eta) = 1$ or $0$ depending on whether they percolate or not.  
If we perform $N$ trials, the number of trials that return $1$ is binomially distributed with mean $R_L(\eta) N$, 
and averaging gives an estimate of $R_L(\eta)$ 
with standard deviation
\begin{equation}
  \label{eq:sigma-R}
  \sigma_{R_L} = \sqrt{\frac{R_L(\eta)\,\big(1-R_L(\eta)\big)}{N}} \, .
\end{equation}
Depending on which kind of wrapping cluster we are looking for, this is roughly $0.4 N^{-1/2}$.  

When we look for the $\eta_L$ where $R_L(\eta)$ crosses $R_\infty$, 
the error on $\eta_L$ is given by
\begin{equation*}
  \sigma_{\eta_L} = \frac{\sigma_{R_L}}{R'_L(\eta_L)} \, . 
\end{equation*}
Since the slope $R'_L(\eta_L)$ grows as $0.361 L^{3/4}$ 
(see Figure~\ref{fig:steigung}) this gives 
\[
 \sigma_{\eta_L} \approx N^{-1/2} \,L^{-3/4} \, . 
\]
These are the error bars shown in Figure~\ref{fig:etac}.

The extrapolated value for $\eta_c$ is computed from simulations
for $D$ different system sizes $L$, which in a weighted linear
regression as in Figure~\ref{fig:etac} yields an error roughly
$\sqrt{D}$ times smaller than the error bars of the underlying data points.  

Finally, we average our estimates of $\eta_c$ from $R_L^e$ and $R_L^b$.  
Assuming that these estimates are only weakly correlated reduces the error bars by 
another factor of $\sqrt{2}$.

The error bars shown in Table~\ref{tab:etas} are the result of 
simulating roughly $D=50$ system sizes ranging from $L=8$
to $L=2048$, with sample sizes $N$ ranging from $10^{10}$
for the systems with $L \le 100$, to $10^9$ for $100 < L \le 500$, 
to $10^6$ for $500 < L \le 2048$.

We ran these simulations in parallel on several computer clusters
with greatly varying computational power.  In total, our simulations
would have taken about 400 years if done only on the laptop on which
this paper was written.

\section{Conclusions}

We have shown that the union-find approach to estimating percolation thresholds 
introduced by Newman and Ziff~\cite{newman:ziff:01} can be applied in the continuous case.  
With the help of an algorithm for estimating $\eta_c$ that runs in essentially linear time 
as a function of the number of objects at criticality, we have obtained new estimates for 
$\eta_c$ in a variety of continuum percolation models that are several orders of magnitude 
more accurate than previous results.  In the process, we have confirmed the predictions of 
conformal field theory for these models, both for the finite-size scaling exponent $\nu$ and the 
probabilities that various kinds of wrapping clusters exist at $\eta_c$ on periodic boundary conditions.  

\begin{acknowledgements}
S.M. thanks the Santa Fe Institute for their hospitality.  C.M. is supported by the 
National Science Foundation through grant CCF-1219117 and by the 
Air Force Office of Scientific Research and the Defense Advanced
Research Projects Agency through grant FA9550-12-1-0432.  We are grateful to Robert Ziff and Mark Newman 
for helpful conversations.
\end{acknowledgements}

\bibliography{percolation,mertens}

\end{document}